\documentclass[journal]{IEEEtran}

\usepackage{xcolor,soul,framed} %,caption

\colorlet{shadecolor}{yellow}
\usepackage[pdftex]{graphicx}
\graphicspath{{../pdf/}{../jpeg/}}
\DeclareGraphicsExtensions{.pdf,.jpeg,.png}

\usepackage[cmex10]{amsmath}
\usepackage{float}
\usepackage{array}
\usepackage{mdwmath}
\usepackage{diagbox}
\usepackage{mdwtab}
\usepackage{eqparbox}
\usepackage{siunitx}
\usepackage{url}

\begin{document}

\title{Unified Noise-aware Network for Low-count PET Denoising}
  \author{Huidong~Xie$^*$,~\IEEEmembership{Student Member,~IEEE,}
      Qiong~Liu$^*$,~\IEEEmembership{Student Member,~IEEE,}
      Bo~Zhou$^*$,~\IEEEmembership{Student Member,~IEEE,}
      Xiongchao~Chen,~\IEEEmembership{Student Member,~IEEE,}
    Xueqi~Guo,~\IEEEmembership{Student Member,~IEEE,}
      and~Chi~Liu,~\IEEEmembership{Senior Member,~IEEE}

% \thanks{This work involved human subjects or animals in its research. The authors confirm that all human/animal subject research procedures and protocols are exempt from review board approval.}
\thanks{This work involved human subjects in its research. The use of human data in this study was approved by the Institutional Review Board (IRB) of University of Bern and Shanghai Ruijin Hospital.}
\thanks{Huidong Xie, Qiong Liu, Bo Zhou, Xiongchao Chen, Xueqi Guo, and Chi Liu are with the Department of Biomedical Engineering. (e-mail: {huidong.xie, qiong.liu, bo.zhou, chi.liu}@yale.edu).}
\thanks{Chi Liu is with the Department of Radiology and Biomedical Imaging at Yale University}
\thanks{Huidong Xie, Qiong Liu, and Bo Zhou contributed equally to this work.}}
%   \thanks{M. Roberg is with TriQuint Semiconductor, 500 West Renner Road Richardson, TX 75080 USA (e-mail: michael.roberg@tqs.com).}% <-this % stops a space
%   \thanks{T. Reveyrand is with the XLIM Laboratory, UMR 7252, University of Limoges, 87060 Limoges, France (e-mail: tibault.reveyrand@xlim.fr).}%
%   \thanks{I. Ramos and Z. Popovic are with the Department of Electrical, Computer and Energy Engineering, University of Colorado, Boulder, CO, 80309-0425 USA (e-mail: ignacio.ramos@colorado.edu; zoya.popovic@colorado.edu).}% <-this % stops a space
%   \thanks{E. Falkenstein is with Qualcomm Inc., 6150 Lookout Road
% Boulder, CO 80301 USA (e-mail: erez.falkenstein@gmail.com).}

% \markboth{IEEE TRANSACTIONS ON MICROWAVE THEORY AND TECHNIQUES, VOL.~60, NO.~12, DECEMBER~2012
% }{Roberg \MakeLowercase{\textit{et al.}}: High-Efficiency Diode and Transistor Rectifiers}

% ====================================================================
\maketitle

% === ABSTRACT ====================================================================
% =================================================================================
\begin{abstract}
As PET imaging is accompanied by substantial radiation exposure and cancer risk, reducing radiation dose in PET scans is an important topic. However, low-count PET scans often suffer from high image noise, which can negatively impact image quality and diagnostic performance. Recent advances in deep learning have shown great potential for recovering underlying signal from noisy counterparts. However, neural networks trained on a specific noise level cannot be easily generalized to other noise levels due to different noise amplitude and variances. To obtain optimal denoised results, we may need to train multiple networks using data with different noise levels. But this approach may be infeasible in reality due to limited data availability. Denoising dynamic PET images presents additional challenge due to tracer decay and continuously changing noise levels across dynamic frames. To address these issues, we propose a Unified Noise-aware Network (UNN) that combines multiple sub-networks with varying denoising power to generate optimal denoised results regardless of the input noise levels. Evaluated using large-scale data from two medical centers with different vendors, presented results showed that the UNN can consistently produce promising denoised results regardless of input noise levels, and demonstrate superior performance over networks trained on single noise level data, especially for extremely low-count data.

\end{abstract}

%To achieve optimal denoising performance for images with different noise levels, one may need to train multiple networks and choose the corresponding network for the best performance. It would be time-consuming in reality to

% === KEYWORDS ====================================================================
% =================================================================================

% \fi
% \IEEEpeerreviewmaketitle

% ====================================================================
% ====================================================================
% ====================================================================

% === I. INTRODUCTION =============================================================
% =================================================================================
\section{Introduction}

\IEEEPARstart{P}{ositron} Emission Tomography (PET) is a functional imaging modality widely used in oncology \cite{bar2000pet,rohren2004clinical}, cardiology \cite{schwaiger2005pet,gould1991pet}, and neurology \cite{edison2007amyloid,clark2012cerebral} studies. PET scans require the injection of a radioactive tracer, with the dosage level carefully guided by the principle of As Low As Reasonably Achievable (ALARA) \cite{prasad2004radiation}. Given the growing concern about radiation exposure to patients and healthcare providers, decreasing the PET injection dose is desirable \cite{robbins2008radiation}. However, PET image quality is negatively affected by the reduced injection dose and may affect diagnostic performance such as the identification of low-contrast lesions \cite{schaefferkoetter2015initial,lu2019investigation}. Therefore, reconstructing high-quality images from noisy input is an important topic.

The commonly utilized iterative reconstruction methods for nuclear imaging such as maximum likelihood expectation maximization (MLEM) \cite{shepp_maximum_1982} and ordered subset expectation maximization (OSEM) \cite{hudson_accelerated_1994} are particularly vulnerable to the effects of Poisson noise when dealing with low-count PET data \cite{vaissier2013fast, bland2017mr}. Modified EM algorithms have been proposed for improved reconstructions \cite{green1990bayesian}. Iterative algorithms incorporating a Total Variation (TV) regularization can be used for noise reduction \cite{wang_low_2014}. However, iterative methods are time-consuming and fail to produce high-quality reconstructions in challenging cases.

Deep learning has emerged as a new class of reconstruction algorithm and has shown successful implementations in various medical imaging tasks \cite{wang_perspective_2016}. In the context of low-count PET denoising, deep learning has been widely explored in the literature. For instance, Xu \textit{et al.} \cite{xu_200x_2017} proposed a U-net-like network \cite{ronneberger_u-net_2015} that synthesizes full-count PET images from their low-count counterparts. Another study \cite{zhou2020supervised} proposed a supervised network with cycle consistency for improved performance. Task-specific perceptual loss \cite{ouyang_ultra-low-dose_2019} can also be used to ensure the consistency of clinical features between low-count input and synthesized full-count images. Moreover, incorporating prior information from Computed Tomography (CT) and Magnetic Resonance Imaging (MRI) within a deep-learning network can improve low-count PET results and achieve unsupervised denoising \cite{cui2019pet}. In addition to these previous studies on static low-count PET, methods have also been developed for simultaneous motion correction and low-count PET reconstruction \cite{zhou2021mdpet,zhou2020simultaneous,zhou2023fast}, and have demonstrated further improved reconstruction quality. Techniques proposed for other imaging modalities, including but not limited to few-view CT \cite{xie_deep_2019, xie_deep_2020, zhou2021dudodr}, low-dose CT \cite{shan_3-d_2018, zhou2022dudoufnet}, few-view SPECT \cite{xie_increasing_2022, ryden_deep_2020, chen2023dudoss}, low-dose SPECT \cite{aghakhan_olia_deep_2022}, can also be easily adapted for low-count PET denoising.

% From another perspective, the unsatisfactory generalizability of deep learning networks highlights the significant influence of the training dataset on the model's performance\cite{liu2021impact}. Denosing networks that are trained on images with higher noise levels tend to exhibit more aggressive denoising. In contrast, those trained on images with lower noise levels tend to preserve the resolution of the image better. A previous study demonstrates that the weighted results from two networks provide better trade-offs between reducing noise and preserving resolution\cite{liu2022personalized}. Nonetheless, when multiple networks are introduced, determining the weighting factor through observation is not practical. The assumption we make is that the weighting factors are associated with the characteristics of image noise level features. Thus, the optimal weighting factors can be learned and constructed in a unified network for multi-low dose PET denoising. 

\begin{figure*}[!t]
\centerline{\includegraphics[width=\textwidth]{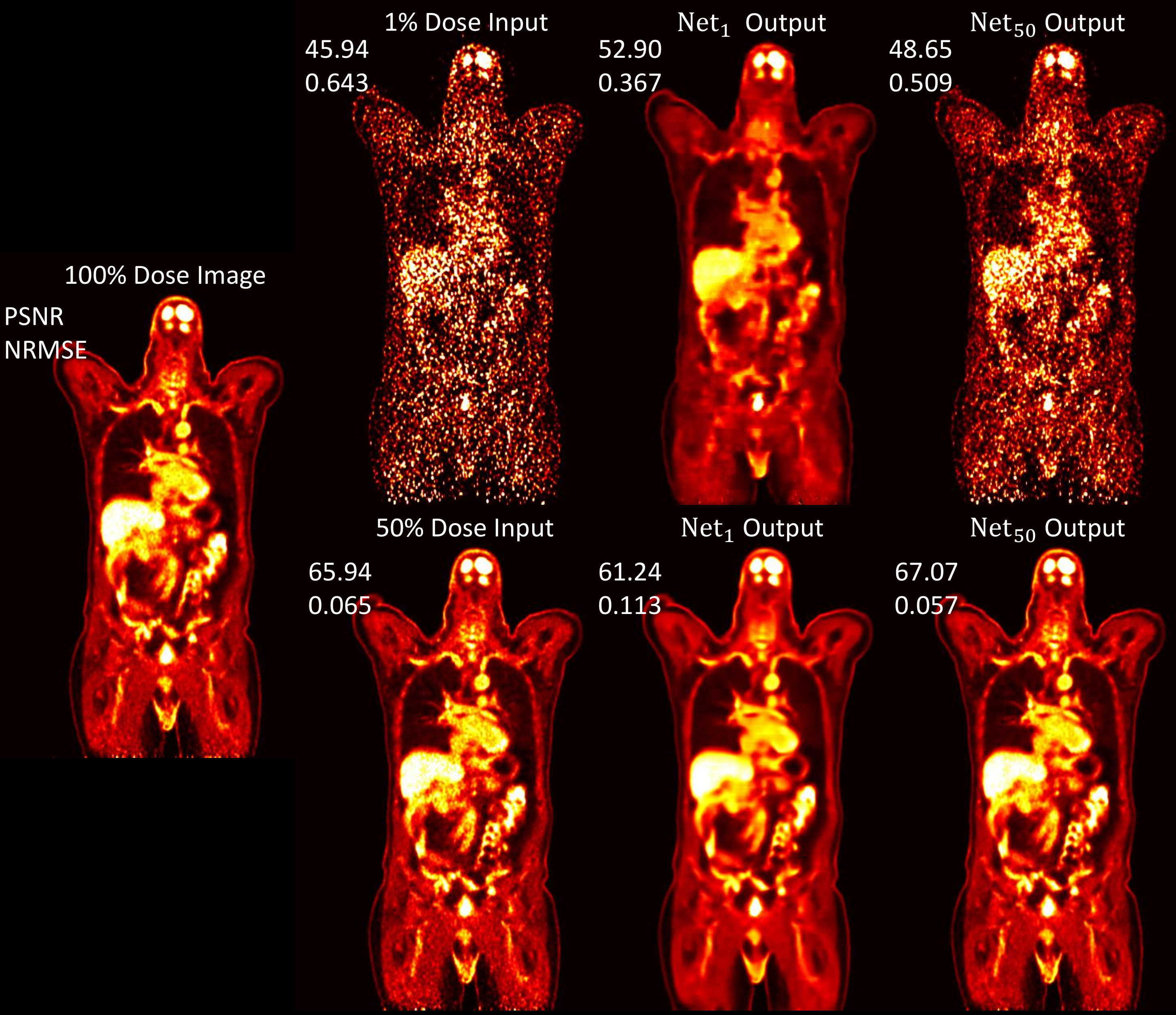}}
\caption{Denoising results generated by networks trained with different count-level inputs. $\mathrm{Net}_1$ and $\mathrm{Net}_{50}$ denote the same networks trained with 1\% and 50\%-count data. PSNR and NRMSE values were computed using the entire 3D volume with the 100\%-dose image as the reference. Low-count images were reconstructed using re-binned listmode data.}
\label{fig1}
\end{figure*}

However, previously proposed methods have limited generalizability to different noise levels. Network trained on images with higher noise levels tend to exhibit more aggressive denoising, while those trained on images with lower noise levels tend to have lower denoising power and preserve the resolution of the image better. For example, a network trained with 1\%-count  PET images may not produce satisfactory results when applied to low-count images with other count levels (e.g., 50\%-count) due to large discrepancies in count levels between the training and testing images. To illustrate this point, we adapted a 3D U-net-like network as an example, and the detailed network structure will be presented later. The networks trained using 1\% or 50\%-count images are denoted as $\mathrm{Net}_1$ or $\mathrm{Net}_{50}$ respectively. Peak signal-to-noise ratio (PSNR) and normalized root mean-squared error (NRMSE) were used as indications of image quality. As shown in Fig. \ref{fig1}, when we used the 1\%-count images as input, $\mathrm{Net}_1$ produced promising denoising results, whereas the $\mathrm{Net}_{50}$ did not have sufficient denoising power and produce low-quality images due to count differences between testing and training data. Similarly, when given a 50\%-count image as input, the $\mathrm{Net}_1$ produced over-smoothed images due to high noise in the training data, resulting in even worse quantitative measurements compared with the 50\%-count input.

One possible solution is to combine data with different noise levels for neural network training. However, our previous work \cite{liu2022personalized,liu_impact_2021} demonstrated that grouping data of different noise levels to train a neural network results in sub-optimal images due to count discrepancies in the training data.

The reconstructions in Fig. \ref{fig1} show that in order to achieve optimal denoised results, it is necessary to train networks using data with the same noise level as the testing data. However, training multiple networks using different noise levels may not be feasible in reality due to limited data availability. Also, in the case of denoising dynamic PET images, the amount of detected photons is continuously decreasing due to tracer decay, resulting in constantly changing noise levels. This problem will be more severe in tracers with short a half-life such as Rubidium-82, which has a 75-second half-life and is widely used for cardiac PET imaging \cite{flotats_hybrid_2011}. 

% A previous study demonstrates that the weighted results from two networks provide better trade-offs between reducing noise and preserving resolution\cite{liu2022personalized}. Nonetheless, when multiple networks are introduced, determining the weighting factor through observation is not practical. The assumption we make is that the weighting factors are associated with the characteristics of image noise level features. Thus, the optimal weighting factors can be learned and constructed in a unified network for multi-low dose PET denoising. 

Given a low-count PET image with 25\%-count level, both $\mathrm{Net}_1$ and $\mathrm{Net}_{50}$ in Fig.\ref{fig1} cannot produce optimal images. A network with denoising power between $\mathrm{Net}_1$ and $\mathrm{Net}_{50}$ is needed for desirable results. Instead of training another network using 25\%-count images (i.e., $\mathrm{Net}_{25}$), if we can combine networks with different denoising powers, we may be able to produce optimal denoising results for images with intermediate count levels. As discussed in our previous work \cite{liu2022personalized}, combining networks with different denoising powers is indeed a possible solution for this problem. The weighted results from the two networks provide better trade-offs between reducing noise and preserving resolution. However, in our previous work \cite{liu2022personalized}, the weights combining different networks need to be manually fine-tuned for the best performance. When multiple denoising networks are introduced, manually determining the weighting factors is not practical. Our assumption is that the weighting factors are associated with the characteristics of image noise level features. Thus, the optimal weighting factors can be learned and constructed in a unified network for multi-low-count PET denoising. 

In this work, we proposed a unified noise-aware network (UNN) for low-count PET denoising. The proposed method generates weights based on the input images to combine networks with different denoising powers and produce the final denoised results. Our results demonstrate that UNN consistently generates promising denoised results for images with any count levels. For extremely low-count data, UNN also outperformed individual networks trained using data with a specific count level (e.g., $\mathrm{Net}_1$ and $\mathrm{Net}_{50}$).

\section{Methods}

\subsection{Multi-institutional Low-count PET data}
The proposed method was evaluated using low-count PET data from two different medical centers in Switzerland and China. The first dataset was collected at the Department of Nuclear Medicine, University of Bern, Bern, Switzerland \cite{xue2021cross}. 209 subjects with the \textsuperscript{18}F-FDG tracer were included in this dataset. All data were acquired using a Siemens Biograph Vision Quadra whole-body PET/CT system. Images were reconstructed using OSEM with 6 iterations and 5 subsets. A $5 \si{mm}$ FWHM Gaussian filter with was applied after reconstructions. The reconstruction matrix size is $644\times 440\times 440$ with a $1.65 \times 1.65 \times 1.65 \si{mm}^3$ voxel size. We randomly selected 99 subjects for training, 10 subjects for validation, and 100 subjects for testing.

The second dataset was collected at the Ruijin Hospital, Shanghai, China \cite{xue2021cross}. 204 subjects with \textsuperscript{18}F-FDG tracer were included in this dataset. All data were acquired using a United Imaging Healthcare uExplorer whole-body PET/CT system. Images were reconstructed using the OSEM algorithm with 4 iterations and 20 subsets. A $5 \si{mm}$ FWHM Gaussian filter was applied after reconstructions. The reconstruction matrix size is $676 \times 360\times 360$ with a $2.88\times 1.66 \times 1.66 \si{mm}^3$ voxel size. We randomly selected 94 subjects for training, 10 subjects for validation, and 100 subjects for testing.

Low-count data were obtained by equally sampling the listmode events into 1\%, 2\%, 5\%, 10\%, 25\%, and 50\% low count levels. Both low-count and full-count reconstructions were provided by using the vendor software.

\subsection{Network Structure}

\begin{figure*}[!t]
\centerline{\includegraphics[width=\textwidth]{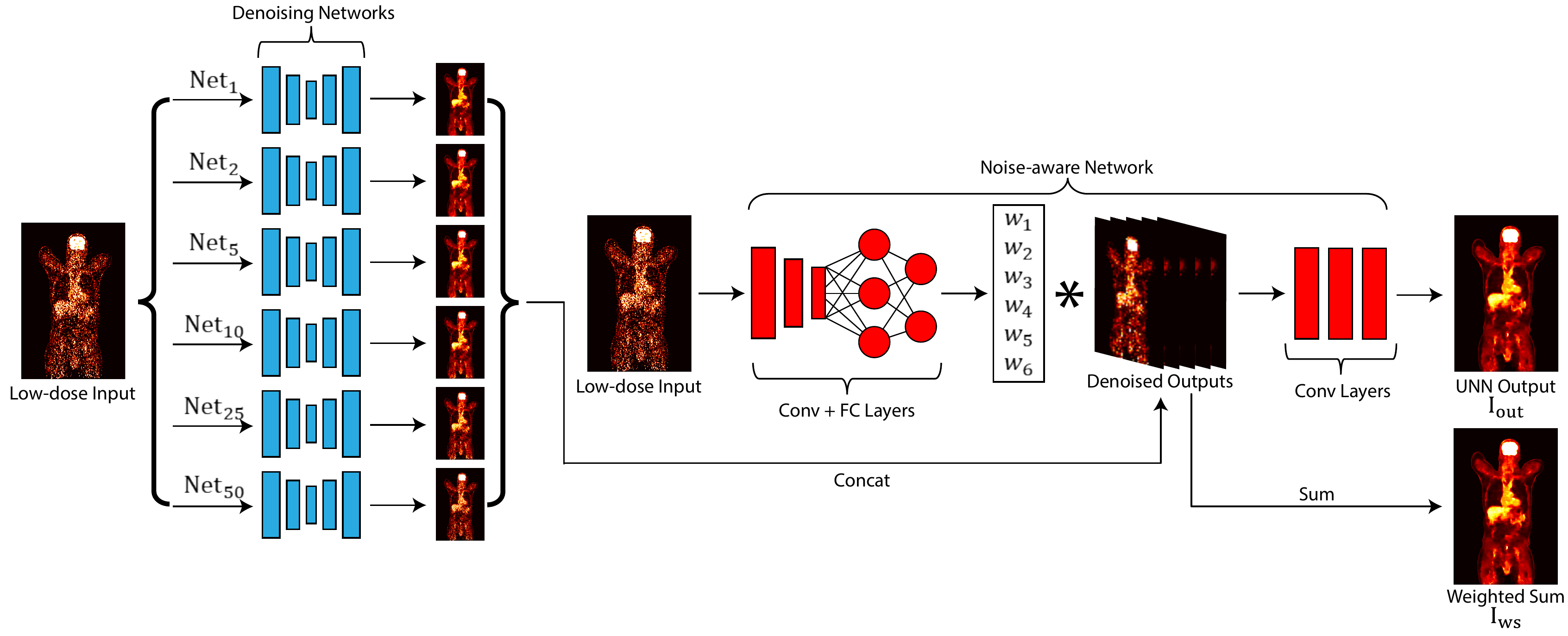}}
\caption{Overall framework of the proposed Unified Noise-aware (UNN) network. It consists of six denoising networks ($\mathrm{Net}_1$ to $\mathrm{Net}_{50}$) and one noise-aware network. $w_1$ to $w_6$ are six personalized weighting parameters to combine denoised outputs from six denoising networks. FC: fully-connected layers. Conv: convolutional layers.}
\label{fig2}
\end{figure*}

The overall structure of the proposed UNN method is presented if Fig. \ref{fig2}. UNN consists of six different supervised denoising networks trained with 1\%, 2\%, 5\%, 10\%, 25\%, and 50\% low-count data respectively and one noise-aware network to generate 6 weighting parameters for the combination of six denoising networks. The six denoising networks are denoted as $\mathrm{Net}_{n}$, where $n$ represents the percentage of photon counts (e.g., $\mathrm{Net}_1$ represents the network trained with 1\%-count data). These six denoising networks share the same network structure with different trainable parameters. These networks take a batch of 3D low-count PET image patches as input and aim to synthesize full-dose counterparts. Each of these networks contains four 3D convolutional layers for down-sampling and another four 3D transpose-convolutional layers for up-sampling. $1\times3\times3$ convolutional kernels were used for both up-sampling and down-sampling layers so that the dimension along the z-axis remains unchanged throughout the network. Zero-padding was not implemented in these layers. To capture features between different slices, a three-layer dense-net \cite{huang_densely_2017} was added after each down-sampling and up-sampling layer. The convolutional kernel size in these dense-net blocks is $5\times3\times3$ with zero-padding so that features on neighboring slices could be seen by the network. Channel and Squeeze-and-Excitation attention blocks \cite{roy_recalibrating_2019} were added after each down-sampling and up-sampling layer, giving important features higher weights during the training process. Rectified Linear Unit (ReLU) is used after all the layers except for the last one.

The noise-aware network aims at generating 6 weighting parameters to combine the outputs of the six denoising networks to generate the final denoised results. The noise-aware network takes a batch of low-count patches as input and generates 6 personalized weighting parameters to combine networks with different denoising powers (i.e., $\mathrm{Net}_1$ to $\mathrm{Net}_{50}$). The noise-aware network consists of five 3D convolutional layers for down-sampling. The convolutional kernel size used in these five layers is $3\times3\times3$ with zero-padding. Strides equal $(1,2,2)$ in the first and the third layers and equal $(2,2,2)$ in other three layers. The down-sampled features are flattened and fed into two fully-connected layers with 18 and 6 neurons, respectively. Rectified Linear Unit (ReLU) is used in all five convolutional layers and the first fully-connected layer. Softmax activation is used after the last fully-connected layer to ensure the sum of all the personalized denoising weights equals to 1. Six weighted denoised outputs are concatenated to generate 6-channel feature maps, each of them represents the weighted denoised output from a denoising network (i.e., $\mathrm{Net}_1$ to $\mathrm{Net}_{50}$).

To generate the final denoised result, these 6-channel feature maps are then fed into another six 3D convolutional layers to squeeze to 1 channel. These six 3D convolutional layers use kernel size $3\times3\times3$ with zero-padding and stride equals to 1. All convolutional layers except the last layers in all denoising and noise-aware networks have 32 filters. The last layers in these networks have 1 filter.

\subsection{Optimization and Training}
The proposed UNN network was trained in a supervised manner using the 100\% full-dose image as the reference. The training process is divided into 2 separate stages. In the first stage, six denoising networks $\mathrm{Net}_1$ to $\mathrm{Net}_{50}$ are individually trained to convergence. In the second stage, these six denoising networks are fixed and the noise-aware network is introduced to obtain the final output.

The objective function used to optimize all the trained networks includes mean-absolute-error (MAE) and SSIM \cite{wang_image_2004}. MAE loss can be formulated as: 

\begin{equation}
\ell_{\mathrm{MAE}}(Y,X)=\frac{1}{N_b W  H D}\sum_{i=1}^{N_b}||Y_i-X_i||_1,  \label{eqn:5}
\end{equation}

where $N_b$ is the input batch size. $W$, $H$, and $D$ are the height, width, and depth of the input/output image patches, respectively. $X$ and $Y$ represent output image patches and the corresponding training label, respectively.

SSIM measures the structural similarity between two images, with a maximum value equals to 1 when two images are identical. In this work, the convolutional window used to calculate SSIM is set to $11\times11$. The SSIM formula is expressed as:

\begin{equation}
\mathrm{SSIM}(Y, X)=\frac{(2\mu_{Y}\mu_{X}+C_1)(2\sigma_{{YX}}+C_2)}{(\mu_{Y}^2+\mu_{X}^2+C_1)(\sigma_{Y}^2+\sigma_{X}^2+C_2)}  \label{eqn:6}
\end{equation}

where $C_1=(0.01\cdot R)^2$ and $C_2=(0.03\cdot R)^2$ are constants to stabilize the ratios when the denominator is too small. $R$ stands for the dynamic range of pixel values. $\mu_{Y}$, $\mu_{X}$, ${\sigma_{Y}}^2$, ${\sigma_{X}}^2$ and $\sigma_{{YX}}$ are the means of $Y$ and $X$, deviations of $Y$ and $X$, and the correlation between $Y$ and $X$ respectively. SSIM loss used to optimize the network parameters is expressed as: $\ell_{\mathrm{SSIM}}=1-SSIM(Y,X)$. Since both the input and output of the network are 3D image patches, we took advantage of that by computing and averaging SSIM values along three different axes. The composite loss function for training the six denoising networks is expressed as:

\begin{equation}
\ell(I,I_{out})= \ell_{\mathrm{MAE}}(I,I_{out}) + \lambda_a * \ell_{\mathrm{SSIM}}(I,I_{out})
\end{equation}

where $\lambda_a$ is hyper-parameters used to balance different loss terms. $I$ and $I_{out}$ represent full-dose ground-truth and the six denoising network outputs.

When training the noise-aware network, the weighted sum of all the outputs from the six denoising networks was also included as part of the objective function. This weighted sum image is obtained by:

\begin{equation}
\begin{split}
    I_{ws} = w_1 * \mathrm{Net}_1(I_{in}) + w_2 * \mathrm{Net}_2(I_{in}) + w_3 * \mathrm{Net}_3(I_{in}) + \\ w_4 * \mathrm{Net}_4(I_{in}) + w_5 * \mathrm{Net}_5(I_{in}) + w_6 * \mathrm{Net}_6(I_{in})
\end{split}
\end{equation}

where $w_1$ to $w_6$ are 6 personalized weighting parameters obtained from the fully-connected layers and used to combine outputs from 6 denoising networks. $I_{in}$ and $I_{mid}$ represent the low-count input PET images and the corresponding weighted sum image. The composite loss function is expressed as follows when training the noise-aware network:

\begin{equation}
\ell = \ell(I,I_{out}) + \ell(I,I_{mid})
\end{equation}

$\lambda_a = 0.6$ was experimentally fine-tuned for the best SSIM measurements in the validation set. The Adam method \cite{DBLP:journals/corr/KingmaB14} was used for optimizing trainable parameters with exponential decay rates $\beta_1=0.9$ and $\beta_2=0.999$. The Xavier method \cite{pmlr-v9-glorot10a} was used for parameter initialization. All bias terms were initialized to 0. Batch size $N_b=15$ when training the six denoising networks, and $N_b=1$ when training the noise-aware network. Patch-based training was implemented to alleviate memory burden. Image volumes were cropped into $20\times64\times64$ when training the six denoising networks. A larger patch size was used when training the noise-aware network so that the network has a better understanding of the global noise features. For both scanners, the patch size was 20 slices when training the noise-aware network (i.e., $20\times440\times440$ for the Siemens Biograph Vision Quadra scanner, and $20\times360\times360$ for the United Imaging uExplorere scanner). During the testing process, since the GPU memory is not sufficient to process the entire image volume directly, a patch-based testing process was also implemented with 20 slices as a testing patch for both scanners. Testing patches were overlapped with stride equals to 10 slices and then took average in the overlapping region to remove inconsistencies at the patch boundary. Two UNNs were trained separately for each dataset.

\section{Results}
\subsection{Comparison with Individually-trained Networks}

\begin{figure*}[!t]
\centerline{\includegraphics[width=\textwidth]{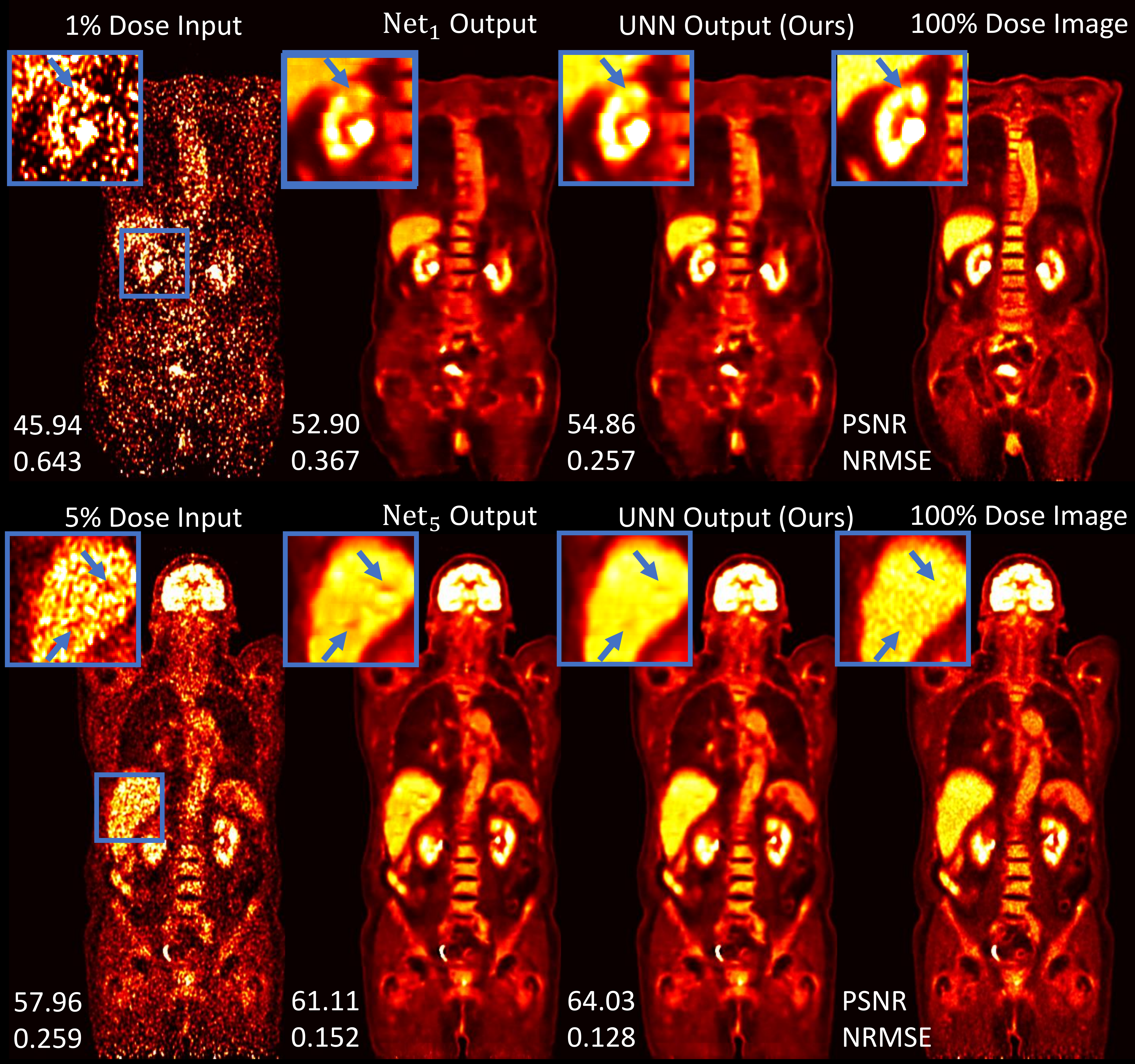}}
\caption{Visual comparison of low-count denoised images generated using different methods. The top and bottom rows are two studies selected from the dataset acquired at the University of Bern. PSNR and NRMSE values were computed using the entire 3D volume with the 100\% dose image as the reference. Blue arrows point to subtle features that are better reconstructed by our proposed method (UNN).}
\label{fig3}
\end{figure*}

To demonstrate the improvements of the proposed UNN method over individually-trained networks trained with different input noise levels ($\mathrm{Net}_1$ to $\mathrm{Net}_{50}$), two sample studies from the dataset acquired at the University of Bern were selected and presented in Fig. \ref{fig3} and the corresponding quantitative measurements are included in Table \ref{table1}. Qualitative and quantitative results show that the proposed UNN consistently produced promising denoising results regardless of input count levels. At lower count levels (1\% to 25\%), the proposed method outperformed individual networks trained with a specific count level. As pointed by the blue arrows in Fig. \ref{fig3}, individually trained denoising networks were unable to recover some subtle details from images with significantly high noise. Combining networks with varying denoising powers, UNN produced images with more uniformed liver with overall fewer artifacts.

\begin{figure*}[!t]
\centerline{\includegraphics[width=\textwidth]{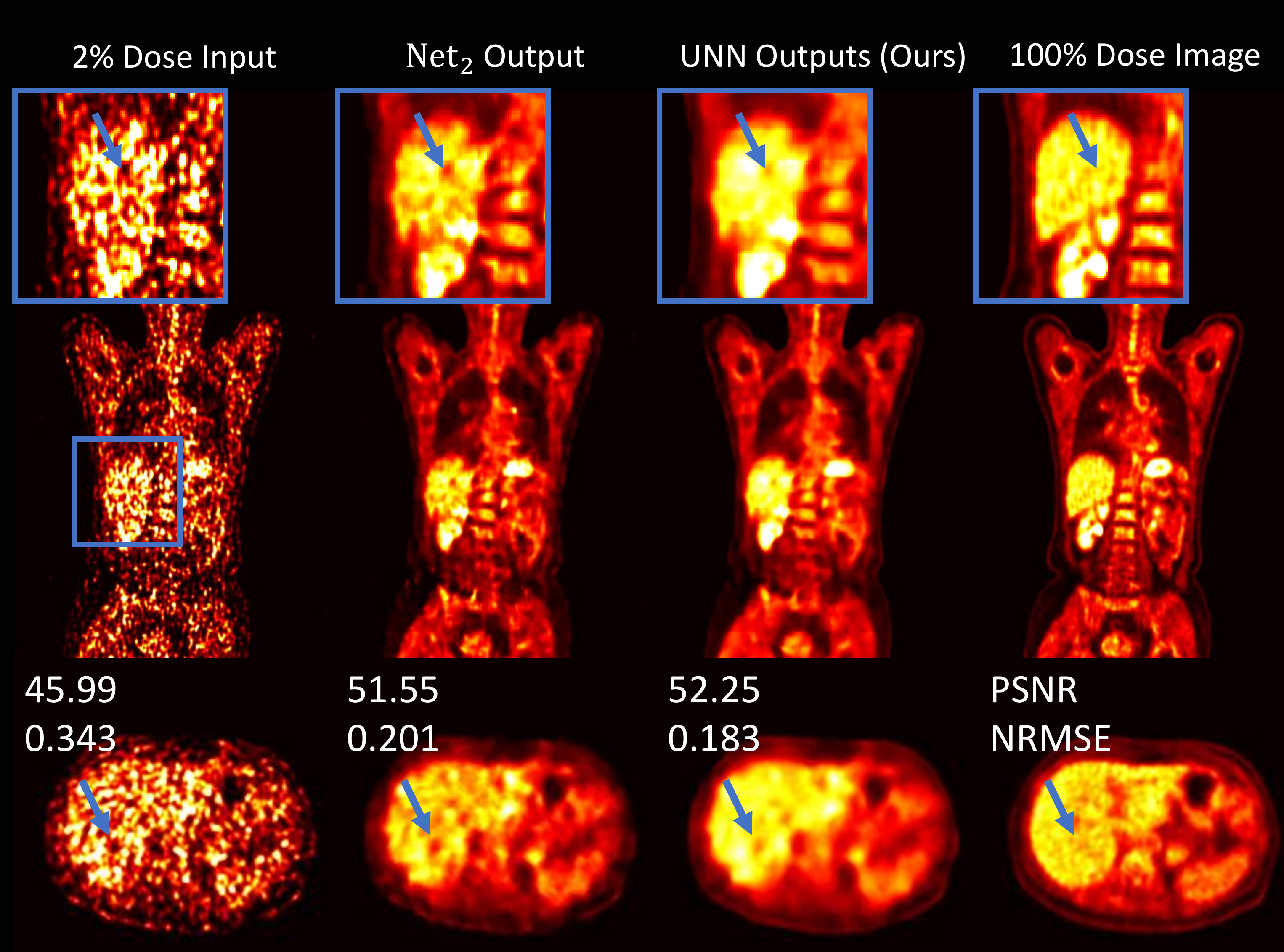}}
\caption{Visual comparison of low-count denoised images generated using different methods. The presented images are from one study selected from the dataset acquired at the Shanghai Ruijin Hospital. PSNR and NRMSE values were computed using the entire 3D volume with the 100\% dose image as the reference. Blue arrows point to subtle features that are better reconstructed by our proposed method (UNN).}
\label{fig4}
\end{figure*}

Another sample study from the dataset acquired at the Shanghai Ruijin Hospital was selected and presented in Fig. \ref{fig4} and the corresponding quantitative measurements are also included in Table \ref{table1}. A similar trend was observed in these images. UNN produced images with overall lower noise and fewer artifacts, leading to a better quantitative measurement. Pointed by the blue arrows in Fig. \ref{fig4}, images reconstructed by UNN have more uniform liver. 

Quantitative results presented in Table \ref{table1} showed that the proposed networks produced promising denoised results for various noise levels and also outperformed individually-trained denoising networks at extremely low count levels (1\% to 25\%). But the improvements became less significant as the count-level increased. At 50\%-count level, combining denoising networks trained at other count-levels did not lead to better performance, and $\mathrm{Net}_{50}$ produced better quantitative measurements over UNN on both datasets. But at 50\%-count, outputs from the individual denoising networks ($\mathrm{Net}_{50}$) and outputs from UNN are visually similar, as presented in Fig. \ref{fig5}.

\subsection{Comparison between $I_{ws}$ and $I_{out}$}

\begin{figure*}[!t]
\centerline{\includegraphics[width=\textwidth]{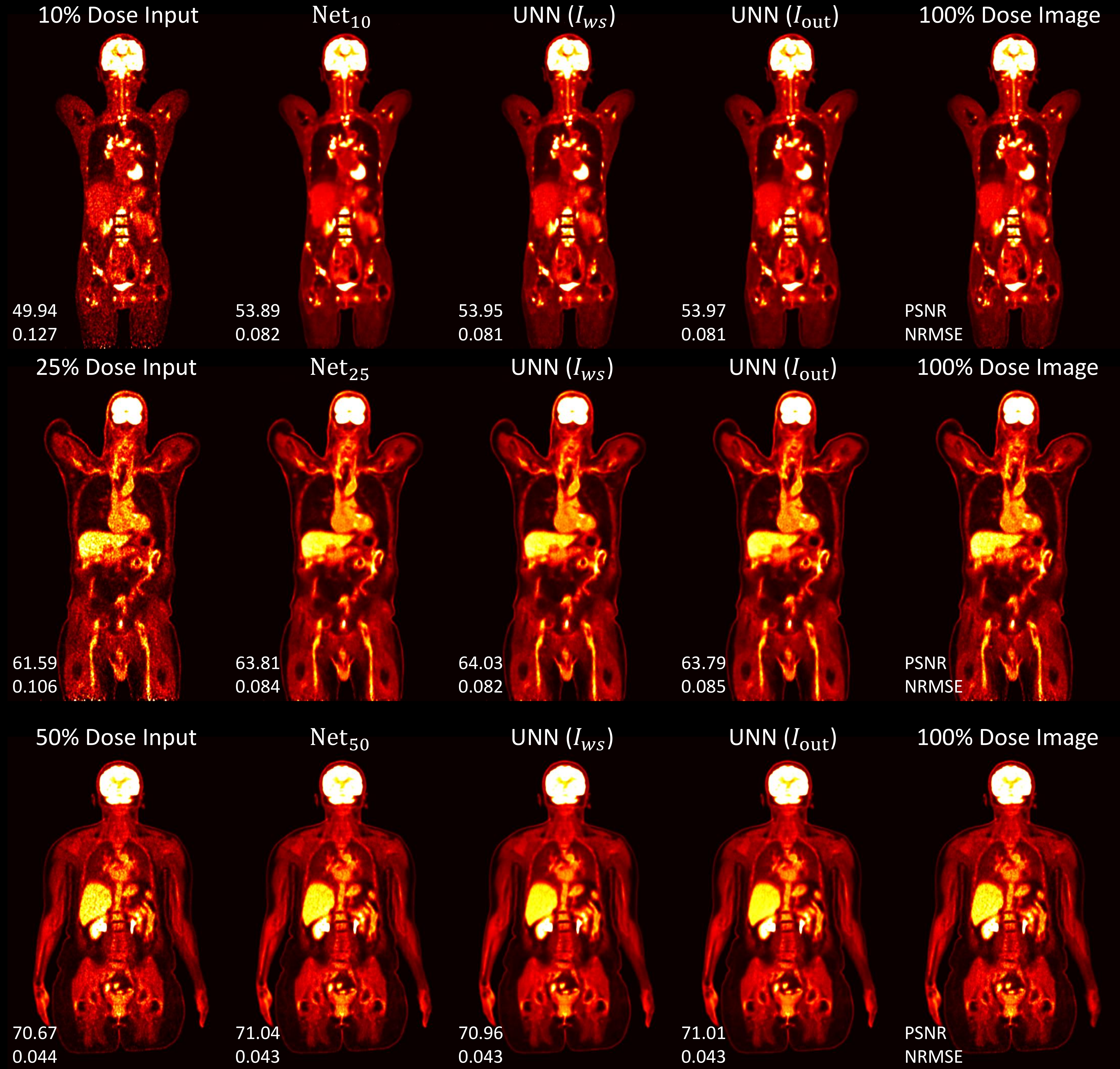}}
\caption{Visual comparison of low-count denoised images reconstructed using different methods from relatively higher count-levels inputs (10\%, 25\%, and 50\%). The presented images are from three studies selected from the dataset acquired at the University of Bern. PSNR and NRMSE values were computed using the entire 3D volume with the 100\% dose image as the reference. At these count-levels, despite slight differences in quantitative measurements, all deep learning methods produced visually similar results. The proposed UNN method consistently produced high-quality low-count images regardless of input count levels.}
\label{fig5}
\end{figure*}

As presented in Table \ref{table1}, using convolutional layers to combine the six denoised outputs ($I_{out}$) led to noticeable improvements over the weighted sum strategy ($I_{ws}$). However, as the count-level increased, the benefits of using convolutional layers at the end of the noise-aware network became less significant. At 10\%, 25\%, and 50\%-count-levels, even though the weighted-sum images led to slightly better quantitative measurements in some cases, the small differences in numbers were not visually noticeable in the reconstructed images. Three sample studies from the dataset acquired at the University of Bern reconstructed with higher count levels (10\%, 25\%, and 50\%) were selected and presented in Fig. \ref{fig5}. At higher count levels, all deep learning methods led to visually promising denoised results. Despite the slight differences in quantitative measurements, there is no noticeable visual difference between the UNN outputs ($I_{ws}$ and $I_{out}$), and the outputs from individually denoising networks ($\mathrm{Net}_{10}$, $\mathrm{Net}_{50}$, and $\mathrm{Net}_{50}$). 

From 1\%-count to 50\%-count, the proposed UNN method consistently produced high-quality denoised results regardless of input count-levels. Individually denoising networks trained with specific count levels are not able to generalize to other count-levels, as demonstrated in Fig. \ref{fig1}.

\subsection{Visualizations of Weighting Parameters}

\begin{figure}[h!]
\centerline{\includegraphics[width=0.5\textwidth]{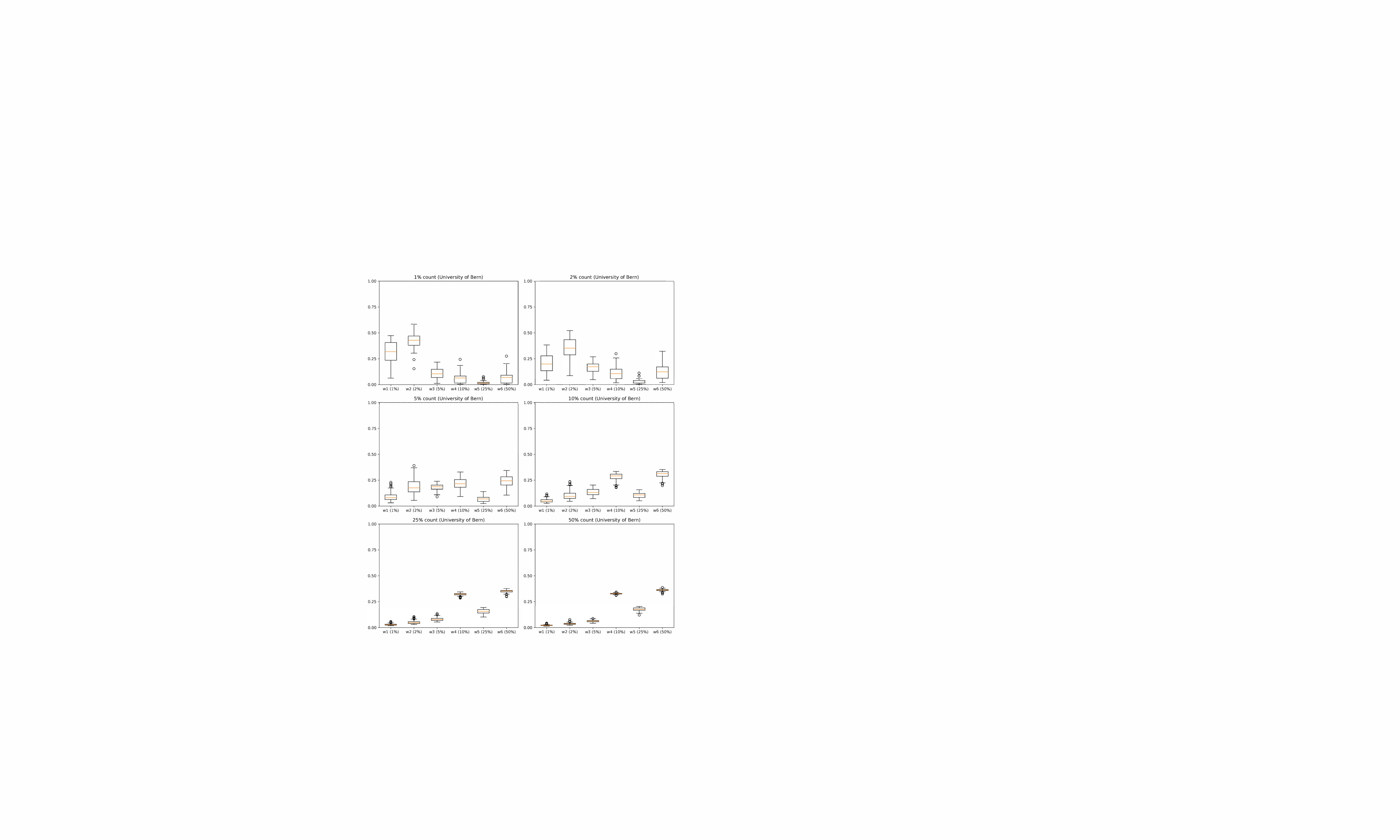}}
\caption{Visualizations of the personalized weighting parameters from the 100 testing patients obtained at the University of Bern. Six plots present the distributions of the weighting parameters at six count levels.}
\label{fig6}
\end{figure}

\begin{figure}[h!]
\centerline{\includegraphics[width=0.5\textwidth]{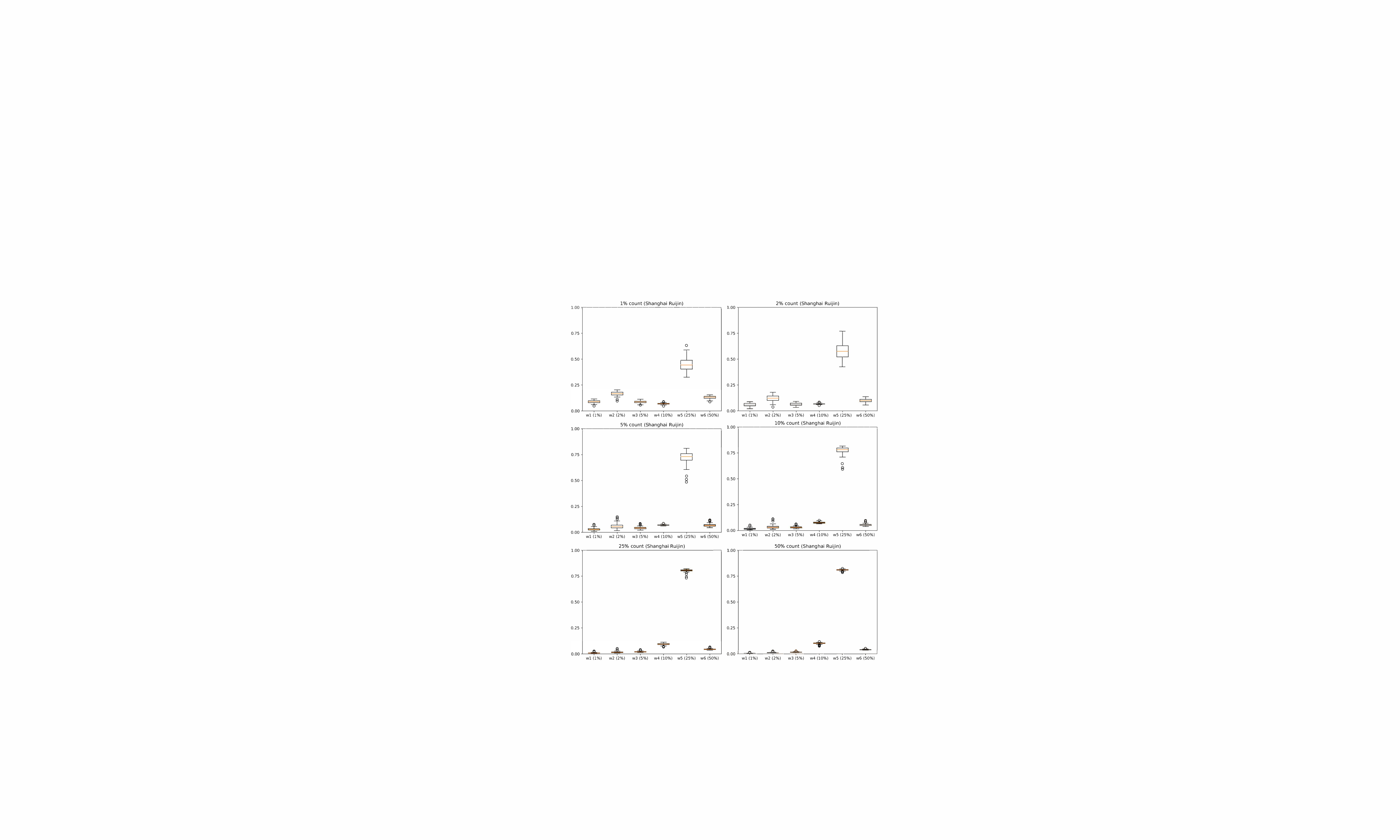}}
\caption{Visualizations of the personalized weighting parameters from the 100 testing patients obtained at the Shanghai Ruijin Hospital. Six plots present the distributions of the weighting parameters at six count levels.}
\label{fig7}
\end{figure}

Figs. \ref{fig6} - \ref{fig7} show the personalized weighting parameters from the testing patients obtained at both institutions. These box plots show that denoising networks with varying denoising powers capture different image features, and the proposed UNN method aims at combining this information for improved reconstructions. Counter-intuitively, the denoising network with the highest weight does not always correspond to the input count level. For example, for the dataset obtained at the University of Bern, at 1\%-count level, images produced by the 2\%-count network ($\mathrm{Net}_2$) contributed the most to the UNN output. For the dataset obtained at the Shanghai Ruijin Hospital, the proposed noise-aware network tends to give the 25\%-count network ($\mathrm{Net}_{25}$) the highest weight across all the count-levels. Note that as the count-level increases, the standard deviations of these personalized weighting parameters decrease. It means that these weighting parameters tend to converge and stabilize high count levels.

\begin{table*}[!h]
\centering
\caption{Quantitative assessment for different methods on two datasets. The best results among different count levels are marked in \textcolor{red}{red}. The measurements were obtained by averaging the values on the testing human studies. The proposed method (UNN) consistently produced promising denoised results regardless of count level. At lower count levels (1\%, 2\%, 5\%, 10\%, and 25\%), the proposed UNN method produced images with better quantification compared with individual denoising networks. "$\dagger$" and "$^*$" indicate statistically significant at $p < 0.0001$ and $p<0.05$, respectively, between the proposed methods (either $I_{ws}$ or $I_{out}$) and the corresponding individually-trained denoising networks at different count levels ($\mathrm{Net}_1$ to $\mathrm{Net}_{50}$).}
\resizebox{\textwidth}{!}{
\begin{tabular}{c|c|c|c|c|c|c}
\hline\hline
\multicolumn{7}{c}{\textbf{Dataset From Shanghai Ruijin Hospital (United Imaging uExplorer Scanner)}}\\
\hline\hline
     \textbf{PSNR/NRMSE} &  1\% Count Input     &   2\% Count Input  &    5\% Count Input  &  10\% Count Input & 25\% Count Input & 50\% Count Input    \\
\hline\
Input  & 45.010 / 0.5706	& 49.010 / 0.3631 & 53.255 / 0.2242 & 56.034 / 0.1638 & 60.123 / 0.1027 & 64.492 / 0.062  \\
\hline
$\mathrm{Net}_1$ & 50.717 / 0.3752	& 53.088 / 0.2539 & 54.323 / 0.2116 & 54.777 / 0.1987 & 55.161 / 0.188 & 55.386 / 0.1827 \\

$\mathrm{Net}_2$ & 50.356 / 0.3630	& 53.885 / 0.2250 & 56.156 / 0.1675 & 57.035 / 0.1502 & 57.889 / 0.1340 & 58.488 / 0.1243  \\

$\mathrm{Net}_5$  & 48.930 / 0.4194 & 53.022 / 0.2475 & 56.494 / 0.1618 & 58.087 / 0.1335 & 59.638 / 0.1094 & 60.646 / 0.0964  \\

$\mathrm{Net}_{10}$  & 49.206 / 0.4152	& 53.122 / 0.2435 & 56.553 / 0.1595 & 58.593 / 0.1258 & 61.064 / 0.0932 & 62.938 / 0.0738 \\

$\mathrm{Net}_{25}$  & 47.563 / 0.4711 & 51.949 / 0.2739 & 55.859 / 0.1708 & 58.268 / 0.1295 & 61.636 / 0.0874 & 64.833 / 0.0597  \\

$\mathrm{Net}_{50}$  & 45.688 / 0.5476 & 50.095 / 0.3308 & 54.417 / 0.2005 & 57.105 / 0.1475 & 61.003 / 0.0941 & \textcolor{red}{64.965 / 0.0592}  \\
\hline
UNN Weighted Sum ($I_{ws}$)  & 48.461 / 0.3688 & 52.743 / 0.2343 & 56.272 / 0.1580 & 58.505 / \textcolor{red}{0.1237$\dagger$} & \textcolor{red}{61.722$\dagger$ / 0.0858$\dagger$} &64.731 / 0.0600  \\
\hline
UNN ($I_{out}$)  & \textcolor{red}{52.303$\dagger$ / 0.2496$\dagger$} & \textcolor{red}{54.685$\dagger$ / 0.1924$\dagger$} & \textcolor{red}{57.021$\dagger$ / 0.1477$\dagger$} & \textcolor{red}{58.600} / 0.1247 & 60.997 / 0.0963 & 63.412 / 0.0744 \\
\hline
\multicolumn{7}{c}{}\\
\hline\hline
\multicolumn{7}{c}{\textbf{Dataset From University of Bern (Siemens Biograph Vision Quadra Scanner)}}\\
\hline\hline
     \textbf{PSNR/NRMSE} &  1\% Count Input     &   2\% Count Input  &    5\% Count Input  &  10\% Count Input & 25\% Count Input & 50\% Count Input    \\

\hline

Input  & 49.237 / 0.4760 & 52.397 / 0.3268 & 56.206 / 0.2095 & 58.851 / 0.1544 & 62.609 / 0.1005 & 66.629 / 0.0640  \\
\hline
$\mathrm{Net}_1$ & 55.778 / 0.2612 & 58.069 / 0.1804 & 59.610 / 0.1455 & 60.304 / 0.1329 & 61.128 / 0.1195 & 61.820 / 0.1096  \\

$\mathrm{Net}_2$ & 55.463 / 0.2662 & 58.152 / 0.1796 & 59.444 / 0.1526 & 60.614 / 0.1294 & 61.625 / 0.1128 & 62.487 / 0.1013  \\

$\mathrm{Net}_5$  & 55.179 / 0.2760 & 57.541 / 0.1926 & 60.566 / 0.1320 & 61.463 / 0.1203 & 63.136 / 0.0963 & 64.465 / 0.0819 \\

$\mathrm{Net}_{10}$  & 55.502 / 0.2560 & 58.268 / 0.1744 & 60.767 / 0.1281 & 62.161 / 0.1085 & 64.045 / 0.0868 & 65.823 / 0.0702  \\

$\mathrm{Net}_{25}$  & 53.305 / 0.3245 & 56.85 / 0.2086 & 60.111 / 0.1377 & 61.938 / 0.1110 & 64.192 / 0.0863 & 66.726 / 0.0636  \\

$\mathrm{Net}_{50}$  & 51.963 / 0.3665 & 55.102 / 0.2456 & 58.664 / 0.1602 & 60.987 / 0.1224 & 64.230 / 0.0846 & \textcolor{red}{67.610 / 0.0577}  \\
\hline
UNN Weighted Sum ($I_{ws}$)  & 56.699 / 0.2055  & 58.587 / 0.1667  & 60.771 / 0.1267 & \textcolor{red}{62.244$^*$} / 0.1059 & \textcolor{red}{64.478$^*$ / 0.0820$^*$} & 66.885 / 0.0621  \\
\hline
UNN ($I_{out}$)  & \textcolor{red}{57.206$\dagger$ / 0.1918$\dagger$} & \textcolor{red}{58.909$\dagger$ / 0.1580$\dagger$} & \textcolor{red}{60.885$\dagger$ / 0.1246$\dagger$} & 62.232 / \textcolor{red}{0.1059$\dagger$} & 64.276 / 0.0836 & 66.718 / 0.0632  \\
\hline

\end{tabular}
}
\label{table1}
\end{table*}

\section{Discussion}
Denoising networks that are trained using images with low count levels tend to exhibit more aggressive denoising and vice versa. Therefore, denoising networks trained with one specific count level do not generalize well to other count levels, as demonstrated in Fig. \ref{fig1}. In this work, we proposed to unify denoising networks with varying denoising power to allow PET image denoising at varying count-levels. By combining different denoising networks, the proposed Unified Noise-aware Network (UNN) produced consistently promising denoised results regardless of the input count-levels. Previous methods proposed for low-count PET denoising only work for one specific count level.

In addition, denoising networks with different denoising powers contain prior information at different count levels. The proposed UNN also produced noticeably better reconstructions at extremely low count levels (1\%, 2\%, and 5\%). As presented in Figs. \ref{fig3} - \ref{fig4} and Table \ref{table1}, UNN produced images with more uniform liver, overall fewer artifacts, and better quantitative measurements. At higher count levels (10\%, 25\%, and 50\%), UNN demonstrated superior generalizability and produced images with similar quality to individual denoising networks.

In UNN, using convolutional layers to combine outputs from individual denoising networks produced overall better results at low count levels (1\%, 2\%, and 5\%). At higher count levels (10\%, 25\%, and 50\%), simple weighted sum combinations outperformed the convolutional strategy. However, the difference between them ($I_{ws}$ and $I_{out}$) is not noticeable visually, as presented in Fig. \ref{fig5}.

As presented in Fig. \ref{fig6}, the standard deviations of the personalized weighting parameters decrease as the count level increases, and we suspect that it may be because of the lower voxel variations in high-count images.

Since all the individual denoising networks ($\mathrm{Net}_1$ to $\mathrm{Net}_{50}$) are independent of each other, they can be run in parallel. Therefore, the inference time of the proposed UNN is comparable to individual denoising networks. The noise-aware network only contains a few additional convolutional and fully-connected layers and does not significantly increase the inference time.

In UNN, a small neural network was used to capture noise information in the input images. In the future, similar techniques may be applicable to other medical imaging problems. For example, for the federated learning problem \cite{zhou_federated_2023}, training a unified neural network to combine local models at different institutions could be a possible approach to address the data privacy concern while improving network performance. In addition, similar to dynamic convolution \cite{xie_segmentation-free_2022}, the unified neural network could be used to produce attention weights to adjust convolutional kernels so that the neural networks can be adapted to different datasets. In the future, we plan to validate the proposed method for dynamic PET imaging, in which the count-level is continuously decreasing over time. 

\section{Conclusion}
In conclusion, we proposed a Unified Noise-aware Network for low-count PET image denoising. UNN aims at unifying denoising models with different denoising powers for improved performance. Previous low-count PET denoising networks are not able to generalize well to other input count levels due to noise level discrepancies between the training and testing images. The proposed UNN method consistently produced promising denoised results for varying input count levels.

\section*{Acknowledgment}
This work was supported by the National Institutes of Health (NIH) grant R01EB025468. Parts of the data used in the preparation of this article were obtained from the University of Bern, Department of Nuclear Medicine and School of Medicine, Ruijin Hospital. As such, the investigators contributed to the design and implementation of DATA and/or provided data but did not participate in the analysis or writing of this report. A complete listing of investigators can be found at: “https://ultra-low-dose-pet.grand-challenge.org/Description/”

All authors declare that they have no known conflicts of interest in terms of competing financial interests or personal relationships that could have an influence or are relevant to the work reported in this paper.

% or


% ============================================
%\appendices
%Appendix one text goes here %\cite{Roberg2010}.

%\section{}

%\section*{Acknowledgment}

%The authors would like to thank D. Root for the loan of the SWAP. The SWAP that can ONLY be usefull in Boulder...

\ifCLASSOPTIONcaptionsoff
  \newpage
\fi



%\begin{thebibliography}{1}


\bibliographystyle{IEEEtran}
%\bibliography{IEEEabrv,Bibliography}
\bibliography{citation}

\vfill


\end{document}